\documentclass[twocolumn,showpacs,preprintnumbers,amsmath,amssymb]{revtex4}
\usepackage{graphicx}% Include figure files
\usepackage{dcolumn}% Align table columns on decimal point
\usepackage{bm}% bold math

\begin{document}
\title{Imaging molecular orbitals with laser-induced electron tunneling spectroscopy}% Force line breaks with \\
\author{ XuanYang Lai$^{1}$, RenPing Sun$^{1}$, ShaoGang Yu$^{1}$, YanLan Wang$^{1}$, Wei Quan$^{1}$}
\author{Andr\'e Staudte$^{2}$}
\email{andre.staudte@nrc-cnrc.gc.ca}
\author{XiaoJun Liu$^{1}$}
\email{xjliu@wipm.ac.cn}
\affiliation{$^{1}$ State Key Laboratory of Magnetic Resonance and Atomic and Molecular Physics, Wuhan Institute of Physics and Mathematics, Innovation Academy for Precision Measurement Science and Technology, Chinese Academy of Sciences, Wuhan 430071, China\\
$^{2}$ Joint Attosecond Science Laboratory, National Research Council and University of Ottawa, Ottawa, Ontario K1A 0R6, Canada
}%

\date{\today}% It is always \today, today,

\begin{abstract}
Photoelectron spectroscopy in intense laser fields has proven to be a powerful tool for providing detailed insights into molecular structure. The ionizing molecular orbital, however, has not been reconstructed from the photoelectron spectra, mainly due to the fact that its phase information can be hardly extracted. In this work, we propose a method to retrieve the phase information of the ionizing molecular orbital with laser-induced electron tunneling spectroscopy.  By analyzing the interference pattern in the photoelectron spectrum, the weighted coefficients and the relative phases of the constituent atomic orbitals for a molecular orbital can be extracted. With this information we reconstruct the highest occupied molecular orbital of N$_2$. Our work provides a reliable and general approach for imaging of molecular orbitals with the photoelectron spectroscopy.
\end{abstract}

\maketitle

The molecular orbital theory has been developed to describe the electronic structure of molecules \cite{Pauling1960} and has been extremely successful in explaining the chemical and physical properties of molecules. During the past decades, great progress has been made in developing imaging methods for molecular orbitals in real space using scanning tunneling microscopy \cite{Mohn2012NatNano,Cocker2016Nature}, as well as in reciprocal space using photoemission spectroscopy \cite{Puschnig2009Science,Frietsch2013RSI,Puschnig2011PRB}. The advance of high-repetition rate intense femtosecond lasers and single electron imaging techniques enabled an alternative imaging scheme \cite{Meckel2008Science,Kubel2019NatComm,Murray2011PRL,Wollenhaupt2005ARPC,Grasbon2001PRA,Holmegaard2010NP},
which can trace the evolution of the molecular orbital with subfemtosecond resolution. In this scheme, which we call laser-induced electron tunneling spectroscopy (LETS), the outermost electron of a molecule can be ionized by tunneling through a barrier formed by the molecular potential and the laser electric field \cite{Keldysh1964JETP,Becker2002AdvAtMolOptPhys}. The ionized electron carries information of the molecular orbitals, which is imprinted in the observable molecular-frame photoelectron angular distributions (PADs). Using this approach the symmetry of the highest occupied molecular orbital (HOMO) of N$_2$ and O$_2$ has been directly observed \cite{Meckel2008Science}. By correlating fragmentation channels with the photoelectron spectra, this approach can be extended to image multiple contributing orbitals separately \cite{Akagi2009science}. Furthermore, various intra- and intercycle interference patterns in the PADs are now well understood and separable \cite{Xie2015PRL}, and can be exploited for reconstructing the ionizing orbital. Recently, LETS has clearly shown its inherent ability to image the real-time evolution of valence electron density \cite{Kubel2019NatComm}. However, until now, the molecular orbital has not been reconstructed with LETS. The main reason is that the phase information of the molecular orbital wavefunction, carried by the tunnel-ionized electron wavepacket, is erased during the detection of the photoelectron amplitudes.

%%%%%%%%%%%%%%%%%%%%%%%%%%%%%%%%%%%%%%%%%%%%
\begin{figure}[b]
\includegraphics[width=2.7in]{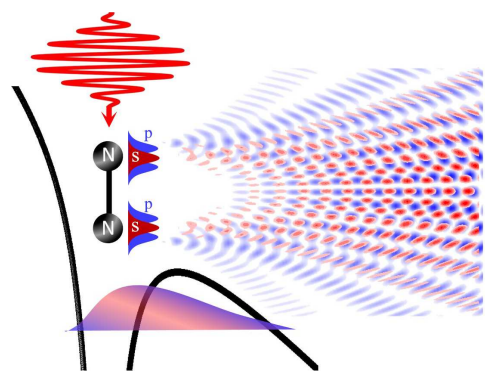}
\caption{(color online) Schematic view of the LETS. The electron of a molecule in HOMO can be ionized by tunneling through a barrier formed by the molecular potential and the laser electric field. According to the LCAO theory, the HOMO of N$_2$ is written as a linear combination of the 2$s$ and 2$p$ atomic orbitals for each core. The tunneling-ionized electron wavepackets from the 2$s$ and 2$p$ atomic orbitals interfere with each other, causing a modulation in the PADs. In addition, the electron wavepackets from the different cores exhibit a two-center interference.}
\label{fig1}%
\end{figure}
%%%%%%%%%%%%%%%%%%%%%%%%%%%%%%%%%%%%%%%%%%%%

In this work, we propose a LETS-based imaging method to completely reconstruct the molecular orbital, by extracting the weighted coefficients and the relative phases of its constituent atomic orbitals. According to molecular orbital theory, the molecular orbitals can be written as a linear combination of atomic orbitals (LCAO) \cite{Huuckel1931ZP,Lennard1929TFS}. For example, for the molecule N$_2$, the HOMO can be approximately written as a simple weighted sum of the 2$s$ and 2$p$ atomic orbitals. Such an approach has been also used for more complex biomolecules, e.g., DNA \cite{Poudel2016PCCP}, using an orthogonalized-LCAO theory. Within our imaging scheme, the tunneling-ionized electron wavepackets from the different constituent atomic orbitals in the framework of LCAO interfere with each other, causing a modulation in the PADs as illustrated in Fig.~\ref{fig1}. By fitting the interference fringes in the PADs, the weighted coefficients and the relative phases of the different constituent atomic orbitals can be extracted and are used to reconstruct the molecular orbital with the corresponding atomic-orbital wavefunctions in terms of the LCAO theory. Here, we experimentally and theoretically identify the fingerprints of the different atomic orbitals in the PADs and show that they can be used to reconstruct the molecular orbital for the example of N$_2$.

It is worth noting that our molecular-orbital imaging scheme is different from the tomographic orbital imaging with high-order harmonic generation (HHG) \cite{Itatani2004Nature,Hassler2010NP,Bertrand2013NP}. The tomographic imaging method requires the measurement of the amplitude and phase of the high-order harmonics \cite{Itatani2004Nature}. Whereas the amplitude of the HHG is easily measurable in experiment, the measurement of the phase is more challenging to access. In practice, the phase information of the high-order harmonics is obtained with theoretical calculations \cite{Itatani2004Nature} or additional measurements \cite{Bertrand2013NP}. In contrast, with our imaging method, the information of the relative phases between the constituent atomic orbitals is imprinted in the interference pattern of the PADs and can be extracted by fitting the interference fringes.

The experiment is performed with cold-target recoil-ion momentum spectroscopy (COLTRIMS) \cite{Ullrich2003RPP,Dorner2000PhysRep} (For our apparatus, refer to our previous publications \cite{Wang2016PRA,Quan2017PRL,Sun2019PRL}). A linearly polarized femtosecond laser pulse is generated from a Ti:sapphire femtosecond laser system with a repetition rate of 5 kHz, a pulse duration of around 30 fs, and a center wavelength of 800 nm. The pulse passes through a $\beta$-BBO crystal to create two-color fields, which are then split into a stretched alignment pulse (800 nm, $\sim$80 fs) and a probe pulse (400 nm, ~35 fs). The two pulses are focused by an on-axis spherical mirror onto a cold supersonic gas jet of N$_2$ molecules inside the vacuum chamber of the COLTRIMS. The probe pulse, which is applied after a time delay, ionizes the molecules aligned by the pre-alignment pulse.  Any alignment angle of the molecule with respect to the polarization of the probe pulse can be achieved by rotating the alignment pulse polarization with a half-wave plate \cite{Meckel2014NP}. The alignment degree  estimated from a 2D angular distribution of ions by using the Coulomb explosion method \cite{Dooley2003PRA,Sun2019PRL,Pavicic2007PRL} is about 0.79 at the time delay  of 4.1 ps. We measure the three-dimensional momenta of the produced photoelectrons in coincidence with the singly charged ions. The alignment pulse creates a low ionization background ($<$0.5 $\%$) compared to the ionization pulse.

%The experiment is performed with cold-target recoil-ion momentum spectroscopy (COLTRIMS) \cite{Ullrich2003RPP,Dorner2000PhysRep} (For our apparatus, refer to our previous publications \cite{Wang2016PRA,Quan2017PRL,Sun2019PRL}). \textcolor{red}{A linearly polarized femtosecond laser pulse from a Ti:sapphire femtosecond laser system passes through a $\beta$-BBO crystal to create two-color fields, which are then split into a stretched alignment pulse (800 nm, $\sim$80 fs) and a probe pulse (400 nm, ~35 fs). The two pulses are focused onto a cold supersonic gas jet of N$_2$ molecules inside the vacuum chamber of the COLTRIMS. The probe pulse ionizes the molecules aligned by the pre-alignment pulse \cite{Meckel2014NP}. The alignment degree estimated by using the Coulomb explosion method \cite{Dooley2003PRA,Sun2019PRL,Pavicic2007PRL} is about 0.79 at the time delay of 4.1 ps. We measure the three-dimensional momenta of the produced photoelectrons in coincidence with the singly charged ions.}

\begin{figure}[tb]
\includegraphics[width=3.5in]{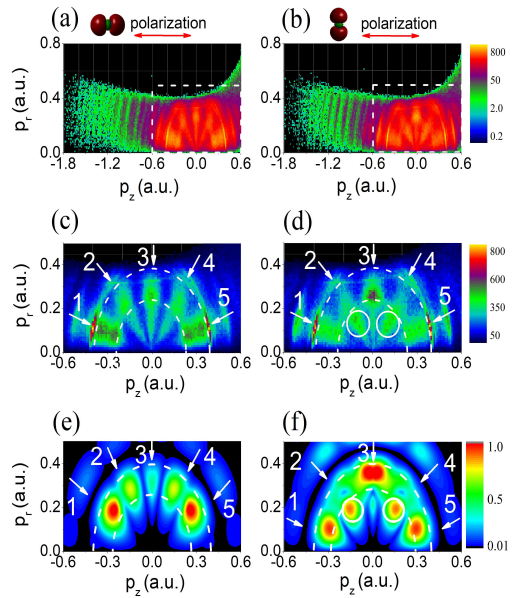}
\caption{(color online) (a) and (b) Experimental results for the parallel and perpendicular alignments of the molecular N$_2$ with respect to the laser polarization, respectively. The sketch above denotes the HOMO of the aligned N$_2$ and the laser polarization direction. $p_z$ and $p_r$ ($p_r=\sqrt{p_x^2+p_y^2}$) represent the momenta parallel and perpendicular to the laser polarization axis, respectively. The data points with $p_r>0.5$ a.u. are removed because of the influence of the spectrometer magnetic field. The laser intensity we used is $I=2 \times {10^{14}}$ W/cm$^2$ with the wavelength of 400 nm,  which is in the tunneling ionization region.} (c) and (d) Enlarged view of the measured PADs in the low-energy region marked by a rectangle in (a) and (b). (e) and (f) Simulated PADs with the M-CQSFA theory.%
\label{fig2}%
\end{figure}

In Figs.~\ref{fig2}(a) and (b), we present the measured PADs for the parallel and perpendicular alignments of the molecular N$_2$ with respect to the laser polarization, respectively. For momenta larger than 0.6 a.u. the PADs show a modulation, which is due to the intercycle interference patterns caused by the well-known above-threshold ionization \cite{Becker2002AdvAtMolOptPhys}. Here, we will focus on the low-energy ($p < 0.6$ a.u.), radial fan-like patterns, which are enlarged in Figs.~\ref{fig2}(c) and (d). There are five radial stripes for both parallel and perpendicular alignments. Interestingly, for the parallel alignment each stripe maximizes in the same energy band, indicated by the dashed, white semi-circles. In contrast, for the perpendicular alignment the second and fourth stripes maximize at lower energies.

To explain the experimental findings we model the corresponding PADs. For an atomic system,  we developed a Coulomb-corrected strong-field approximation (SFA) theory, coined as Coulomb quantum-orbit strong-field approximation (CQSFA) theory \cite{Lai2015PRA,Carla2020RPP}, to reproduce the fan-like patterns in the PADs \cite{Maharjan2006JPB,Shvetsov-Shilovski2016PRA}. Here, we extend our previous work and develop a molecular CQSFA (M-CQSFA), to help understand the alignment-dependent fan-like structures in our measurement. Briefly, the M-CQSFA theory is derived from the exact transition amplitude of an electron from a bound state $|\psi _0 \rangle$ of a molecule to  the continuum state $|\psi_\textbf{p}(t)\rangle$ with drift momentum $\textbf{p}$ \cite{Becker2002AdvAtMolOptPhys}:
\begin{equation}\label{Mpdir}
M(\textbf{p})=-i \lim_{t\rightarrow \infty} \int_{-\infty }^{t }d
t^{\prime}\left\langle \psi_{\textbf{p}}(t)
|U(t,t^{\prime})H_I(t^{\prime})| \psi _0(t^{\prime})\right\rangle \,
\end{equation}
where $H_I(t)=\textbf{r}\cdot \textbf{E}(t)$ is the interaction of a laser field with the electron and $U(t,t^{\prime})$ is the time-evolution operator. In the framework of the LCAO theory \cite{Lennard1929TFS,Huuckel1931ZP}, the initial state $\psi _0$ can be written as a weighted sum of the constituent atomic orbitals of each core. Furthermore, by employing the Feynman path integral formalism and the saddle-point approximation \cite{Lai2015PRA,Carla2020RPP}, the  transition amplitude is written as:
\begin{eqnarray}\label{Mp}
M(\textbf{p}) & \propto &  -
 i \lim_{t\rightarrow \infty }
 \sum_{s} \bigg\{\det \bigg[  \frac{\partial\mathbf{q}_s(t)}{\partial
\mathbf{r}_s(t_{0,s})} \bigg] \bigg\}^{-1/2} e^{i
S(\textbf{q}_s,\textbf{r}_s,t_{0,s},t))} \nonumber \\ & &
\times \mathcal{C}(t_{0,s}) \sum_{a} \left\langle
\mathbf{q}_s(t_{0,s})+\mathbf{A}(t_{0,s})\right. |\textbf{r}\cdot \textbf{E}(t_{0,s})  | \psi_a \rangle  \nonumber \\ & &
\times
{c_a}  \left[e^{i \textbf{p} \cdot \textbf{R}/2}+ (-1)^{l_a} e^{-i \textbf{p} \cdot \textbf{R}/2}\right]
, \,
\end{eqnarray}
where the index $a$ denotes the different atomic orbitals, the index $s$ represents the different quantum trajectories with the action $S(\textbf{q}_s,\textbf{r}_s,t_{0,s},t)$, and $l_a$ is the orbital quantum number of the atomic orbitals. In comparison with the CQSFA theory for atoms \cite{Lai2015PRA}, there is an additional term:  $ e^{i \textbf{p} \cdot \textbf{R}/2}+ (-1)^{l_a} e^{-i \textbf{p} \cdot \textbf{R}/2} $, corresponding to the two-center interference. In practice, the coefficients of the molecular orbital are obtained from quantum chemistry code \cite{Gauss09} and we consider only the ionization of the HOMO of N$_2$, because its contribution is usually much larger than other molecular orbitals \cite{Petretti2010PRL}. In addition, for comparison to the experimental data, the focal-averaged effect of the laser field and the alignment degree of molecular axis are considered in our simulations.

Figs.~\ref{fig2}(e)~and~(f) show the M-CQSFA-simulated PADs of the N$_2$ molecule under parallel and perpendicular alignments, respectively. In agreement with the experiment, our simulations also produce five radial stripes for both parallel and perpendicular alignments. In addition, for perpendicular alignment the second and fourth radial stripes also shift to the low-energy band of the PAD [see the full white circles in Fig.~\ref{fig2}(f)].

\begin{figure}[tb]
\includegraphics[width=3.5 in]{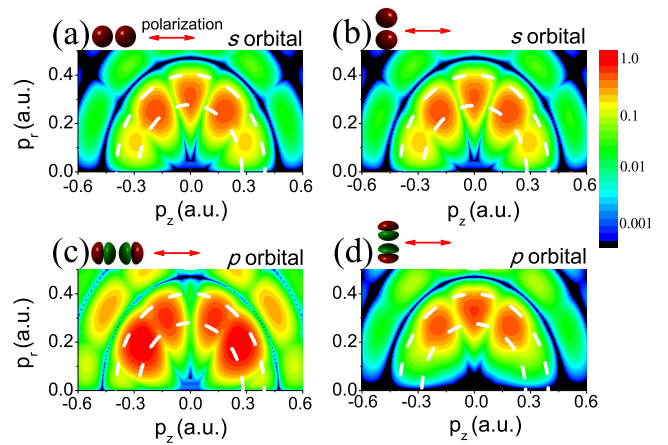}
\caption{ M-CQSFA simulated PADs from the $s$ orbitals [(a) and (b)] and the $p$ orbitals [(c) and (d)] in the HOMO of N$_2$. Left column: parallel alignment; Right column: perpendicular alignment.}%
  \label{fig3}%
\end{figure}

Next, we use our simulation to explain the observed alignment dependence of the PAD. Fig.~\ref{fig3} shows the PADs from the $s$ and $p$ orbitals in the HOMO of N$_2$ for the parallel and perpendicular alignments, respectively.  We find that the low-energy shift of the second and fourth radial stripes shown in the total spectrum of Figs.~\ref{fig2}(f) cannot be observed in Figs.~\ref{fig3}(b)~or~(d). This indicates that the low-energy shift of the radial stripes originates from the interference of the $s$ and $p$ orbital tunneling current.

%%%%%%%%%%%%%%%%%%%%%%%%%%%
%%%%%%%%%%%%%%%%%%%%%%%%%%%

To further uncover the underlying physics of the low-energy shift of the radial stripes, we now analyze the relative phase between the transition amplitudes of the $s$ and $p$ orbitals in Eq.~(\ref{Mp}) (for more details, see Supplemental materials). By comparing the terms relevant to the atomic orbitals, we find that, for the perpendicular alignment, the relative phase is closely related to the combination coefficients $c_a$ of the atomic orbitals in the LCAO theory. A calculation using a standard quantum chemistry package \cite{Gauss09} shows that the combination coefficients for 2$s$ and 2$p$ are opposite in their signs, corresponding to a relative phase of $\pi$. Thus, in the total spectrum the contributions of the $s$ and $p$ orbitals interfere destructively.

Figs.~\ref{fig3}(b) and (d) show that the amplitudes of the second and fourth radial stripes between the two white dashed semicircles are almost the same for the $s$ and $p$ orbitals. Therefore, the relative phase of $\pi$ will lead to a significant ionization suppression in the corresponding region of the total PAD. In the low-energy part of the second and fourth radial stripes the $s$ orbital has a higher probability than the $p$ orbital.  Thus, in the coherent sum of the photocurrent amplitudes from the $s$- and the $p$-orbitals these two radial maxima will appear shifted to lower energy. In contrast, for the first, third, and fifth stripes, the relevant amplitudes of the $s$ and $p$ orbitals differ significantly, resulting in a weak interference between them. Accordingly, these radial stripes retain their maxima in the higher energy region.

For the parallel alignment in Figs.~\ref{fig3}(a) and (c) our analysis shows that there is an additional term relevant to the initial momentum along the laser polarization in the transition amplitude of the $s$ orbitals (for more details, see Supplemental materials). According to the saddle-point equations \cite{Becker2002AdvAtMolOptPhys}, the initial photoelectron momentum along the laser polarization is purely imaginary and thus, the relative phase of the transition amplitudes of the $s$ and $p$ orbitals becomes $\pi/2$. In this case, the total spectrum is approximately equal to the direct sum of the contributions from the $s$ and $p$ orbitals shown in Figs.~\ref{fig3}(a) and (c).

In the following, we will show that the specific low-energy shift of the radial stripes due to the destructive interference provides a LETS-based imaging method to reconstruct the molecular orbital in the framework of the LCAO theory. According to the LCAO theory \cite{Lennard1929TFS,Huuckel1931ZP}, the molecular orbital can be written as a weighted sum of the constituent atomic orbitals of each core.  For the nitrogen atom, there are three bound atomic orbitals: 1$s$, 2$s$ and 2$p$. Thus, the molecular orbital of N$_2$ can be generally written as
\begin{equation}\label{LCAOn2}
\psi_{0}(\textbf{r})\approx \sum\limits_{a=1s,2s,2p} {{|c_a|} e^{i\theta_a}\left[ {\psi _a\left( {{\bf{r}} + {{{{\bf{R}}}} \mathord{\left/
 {\vphantom {{{{\bf{R}}}} 2}} \right.
 \kern-\nulldelimiterspace} 2}} \right) + { s_{a}    }\psi _a\left( {{\bf{r}} - {{{{\bf{R}}}} \mathord{\left/
 {\vphantom {{{{\bf{R}}}} 2}} \right.
 \kern-\nulldelimiterspace} 2}} \right)} \right]},
\end{equation} %
where $|c_a|$ is the weighted coefficients with the phase of $\theta_a$ and the coefficient $s_a$ is relevant to the symmetry of the molecular orbital. Recently, we have proposed and demonstrated a tomographic method to extract the value of $\textbf{R}$ from the measured photoelectron momentum spectra of N$_2$ \cite{Sun2019PRL}. In this work, the combination coefficients ($|c_a|$, $\theta_a$, $s_{a}$) for the molecular orbital are extracted by fitting the interference patterns in the PAD, while the wavefunction of single nitrogen atom is calculated with quantum chemistry code.

First of all, one needs to obtain the coefficient $s_{a}$ from the data. The coefficient $s_{a}=(-1)^{l_a}$ for g symmetry of the molecular orbital, while $s_{a}=(-1)^{l_a+1}$ for u symmetry. If the molecular orbital is in u symmetry, the ionization from the $s$ orbital of the two cores will have a two-center interference term of $\sin(\textbf{p}\cdot\textbf{R}/2)$ \cite{Busuladzic2008PRL}, which leads to the ionization suppression for the photoelectron with $\textbf{p}\perp\textbf{R}$. On the other hand, for the $p$ orbital along the molecular axis, due to its nodal structure, the ionization for the photoelectron with $\textbf{p}\perp\textbf{R}$ is also suppressed. Thus, in the total PAD the ionization distribution of the photoelectron with $\textbf{p}\perp\textbf{R}$ will be vanishing. However, our measurement in Figs.~\ref{fig2}(c) and (d) shows that there are clear ionization distributions for the photoelectron with $\textbf{p}\perp\textbf{R}$. This indicates that the HOMO of N$_2$ should have the g symmetry and therefore, $s_{a}=(-1)^{l_a}$.

%\deleted{It has been shown that the symmetry affects the two-center interference term of the atomic orbital. For the g symmetry, the two-center interference term of the $s$ orbital is $\cos(\textbf{p}\cdot\textbf{R}/2)$, while for the u symmetry, it becomes $\sin(\textbf{p}\cdot\textbf{R}/2)$. For $\sin(\textbf{p}\cdot\textbf{R}/2)$, it leads to the ionization suppression of the $s$ orbital for the photoelectron with $\textbf{p}\perp\textbf{R}$.}

%%%%%%%%%%%%%%%%%%%%%%%%%%%
\begin{figure}[tb]
\includegraphics[width=3.5 in]{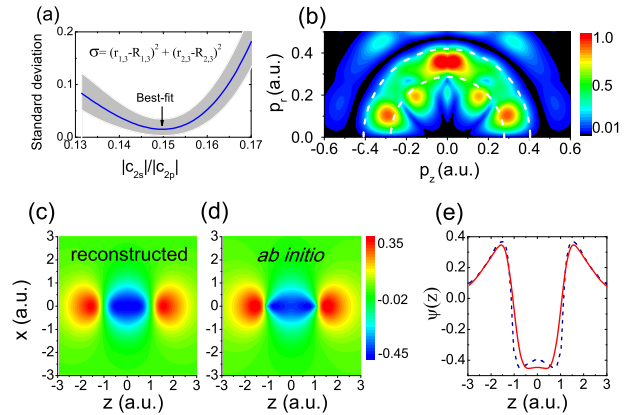}
\caption{(color online) (a) The standard deviation of the relative amplitudes of the different interference stripes between the measurement and the simulation as a function of the ratio of $|c_{2s}|/|c_{2p}|$. The experimental error is shown as shaded area surrounding the solid line. (b) M-CQSFA-simulated PAD for the perpendicular alignment with the best-fit ratio. (c) The reconstructed molecular orbital wavefunction of N$_2$ with the best-fit ratio. (d) The shape of the molecular orbital from an \emph{ab initio} calculation. (e) Cuts along the molecular axis for the reconstructed (solid) and ab initio (dashed) wavefunctions.
}%
  \label{fig4}%
\end{figure}
%%%%%%%%%%%%%%%%%%%%%%%%%%%

Next, we show how to extract the weighted coefficients $|c_a|$ and the phases $\theta_a$ of the different atomic orbitals from the measurements, by fitting the PAD for the perpendicular alignment with our M-CQSFA theory. Usually, such fitting procedure is time-consuming for many coefficients. However, according to the LCAO theory, the combination coefficients are positive or negative real numbers, i.e., the phase $\theta_a$ is 0 or $\pi$. This will greatly simplify our fitting procedure. Additionally, because the $1s$ atomic orbital is tightly bound, its ionization amplitude can be neglected in our case. For simplicity, we set $c_{1s} = 0$ in our calculation. Thus, we consider only the linear combination of the $2s$ and $2p$ atomic orbitals, and vary the ratio of $|c_{2s}|/|c_{2p}|$ and their relative phase in our fitting procedure. Our simulation result shows that, when the relative phase is $\pi$, the low-energy shift of the second and fourth radial stripes in the measurement can be approximately reproduced. To get the best-fit ratio of $|c_{2s}|/|c_{2p}|$, we consider the relative amplitude of the five radial stripes of the fan-like patterns. More specifically, we calculate the amplitude ratio between the first and third radial stripes $r_{1,3} = A_1/A_3$ and the ratio between the second and third radial stripes $r_{2,3} = A_2/A_3$, where $A_i$ ($i$ = 1, 2, and 3) denotes the integrated ionization amplitude of the different radial stripe. By comparing them with the corresponding ratios ($R_{1,3}$ and $R_{2,3}$) obtained from the measured distribution, we get the standard deviation: $\sigma = (r_{1,3}-R_{1,3})^2 + (r_{2,3}-R_{2,3})^2$. Fig.~\ref{fig4}(a) presents the standard deviation as a function of the ratio of $|c_{2s}|/|c_{2p}|$. It clearly shows that the best-fit ratio is about 0.1495. In Fig.~\ref{fig4}(b), we show the M-CQSFA-simulated PAD for the perpendicular alignment with the best-fit ratio, which agrees well with the measurement in Fig.~\ref{fig2}(d) and the simulation in  Fig.~\ref{fig2}(f).

In Fig.~\ref{fig4}(c), we present the reconstructed electron wavefunction of HOMO of N$_2$ according to Eq.~(\ref{LCAOn2}) with the best-fit ratio of 0.1495 and a relative phase of $\pi$ between $2s$ and $2p$. One can see that the reconstructed wavefunction is well consistent with the \emph{ab initio} orbital calculation of the N$_2$ shown in Fig.~\ref{fig4}(d). The minor difference of the electron wavefunction near the two cores ($z=\pm R/2\sim\pm1$ a.u.) [see Fig.~\ref{fig4}(e)] is due to the neglect of the tightly bound $1s$ atomic orbital in our fitting algorithm. A similar difference in the reconstructed wavefunction, relevant to the $1s$ atomic orbital, is also present in the molecular-orbital tomography with HHG \cite{Itatani2004Nature}.

In summary, we have proposed and experimentally demonstrated a method to reconstruct the molecular orbital in the framework of LCAO based on the laser-induced electron tunneling spectroscopy. Using the example of N$_2$ molecule, we have observed a specific interference pattern in PADs under perpendicular molecular alignment, which is used to successfully extract the weighted coefficients and the relative phase of the constituent atomic orbitals for the highest occupied molecular orbital of N$_2$. For larger molecules, the strong-field ionization will lead to the appearance of more relevant peculiar features in the PADs (an example can be found in the Supplemental materials), which will benefit the accurate reconstruction of more complicated molecular orbital with our imaging method. In addition, if combined with a conventional pump-probe scheme, it should be possible to detect the evolution of molecular orbitals in real time and to observe electronic rearrangement during a photochemical reaction. The derived information of the constituting atomic orbitals will provide a comprehensive understanding of the molecular dynamics.

This work is supported by the National Key Program for S$\&$T Research and Development (Grants No. 2019YFA0307702), the National Natural Science Foundation of China (Nos. 11834015, 11922413, 12121004 and 12274420), the Strategic Priority Research Program of the Chinese Academy of Sciences (No. XDB21010400), CAS Project for Young Scientists in Basic Research, Grant No.YSBR-055, the Science and Technology Department of Hubei Province (Grant Nos. 2020CFA029 and 2021CFA078) and K.C. Wong Education Foundation.

X.Y.L., R.P.S. and S.G.Y. contributed equally to this work.

\end{document}